\documentclass[%
reprint,
superscriptaddress,
 amsmath,amssymb,
 aps,
 prl,
showkeys,
]{revtex4-2}

\usepackage{hyperref}
\usepackage{times}

\usepackage{graphicx}% Include figure files
\usepackage{bm}% bold math
\usepackage{xcolor}
\usepackage{cleveref}
\usepackage{marginnote} % to be removed in final version

\newcommand{\real}{\mathbb{R}}

\newcommand{\vF}{\bm{F}}

\newcommand{\vx}{\bm{x}}

\newcommand{\vk}{\bm{k}}
\newcommand{\ve}{\bm{e}}
\newcommand{\vy}{\bm{y}}

\newcommand{\vu}{\bm{u}}

\newcommand{\vzero}{\bm{0}}

\newcommand{\grad}{\nabla}

\newcommand{\mK}{\mathsf{K}}
\newcommand{\mI}{\mathsf{I}}

\begin{document}

\preprint{APS/123-QED}

\title{Wave-driven assembly of quasiperiodic patterns of particles}
\author{Elena Cherkaev}
\affiliation{Department of Mathematics, University of Utah, Salt Lake City UT 84112, USA}
\author{Fernando Guevara Vasquez}
\affiliation{Department of Mathematics, University of Utah, Salt Lake City UT 84112, USA}
\email{fguevara@math.utah.edu}
\author{China Mauck}
\affiliation{Department of Mathematics, University of Utah, Salt Lake City UT 84112, USA}
\author{Milo Prisbrey}
\affiliation{Department of Mechanical Engineering, University of Utah, Salt Lake City UT 84112, USA}
\affiliation{Currently: Acoustics and sensors team, Materials Physics and Applications Group (MPA-11), Los Alamos National Laboratory, Los Alamos NM 87545, USA}
\author{Bart Raeymaekers}
\affiliation{Department of Mechanical Engineering, University of Utah, Salt Lake City UT 84112, USA}
\date{November 2020}

\begin{abstract} % 600 characters maximum
We theoretically show that a superposition of plane waves causes small (compared to the wavelength) particles dispersed in a fluid to assemble in quasiperiodic two or three dimensional patterns. We experimentally demonstrate this theory by using ultrasound waves to assemble quasiperiodic patterns of carbon nanoparticles in water using an octagonal arrangement of ultrasound transducers, and we document good agreement between theory and experiments. The theory also applies to obtaining quasiperiodic patterns in other situations where particles move with linear waves, such as optical lattices. % 597 chars
\end{abstract}
\maketitle
%%%%%%%%%%%%%%%%%%%%%%%%%%%%%%%%%%%%%%%%%%%%%%%%%%%%%%%%%%%%%%%%%%%%%%%%
%\section{Introduction}

We demonstrate that a superposition (finite linear combination) of plane waves assembles quasiperiodic patterns of particles dispersed in a fluid, which is useful to fabricate quasicrystal-like structures \cite{Schechtman:1984:MPL,Yamamoto:1996:CQC,Janot:2012:Q,Stadnik:1999:PPQ} with feature size of approximately one wavelength. Experimental evidence of quasiperiodic patterns of particles or atoms obtained with specific fields, such as lasers \cite{Roichman:2005:HAQ,Wang:2006:ROT,Mikhael:2008:ALT,Sun:2015:FTF} and ultrasound waves \cite{Espinosa:1993:AQC} exists. In contrast, we theoretically derive a general method to obtain prescribed quasiperiodic symmetries (e.g., $8-$fold, $10-$fold in 2D and icosahedral in 3D, among others), for any linear wave-like phenomenon. 
Quasicrystals can exhibit unusual physical properties, e.g. diamagnetic properties, or very low electric conductivity that may be strongly dependent on the temperature, see e.g. \cite{Stadnik:1999:PPQ,Macia:2019:CBP}.
We emphasize that quasicrystals also arise naturally in alloys (e.g. \cite{Ashkin:1987:OTM}) and when combining molecules (e.g. mycelles in \cite{Iacovella:2011:SAS}). However, the specific symmetries are limited by the metals or molecules used, unlike in the theory we demonstrate in this paper.

Quasicrystals are characterized by diffraction patterns with symmetries that do not correspond to any crystalline (periodic) materials, e.g. $10-$fold symmetry in two dimensions \cite{Penrose:1979:PCN,deBruijn:1981:ATP}.
Mathematically they can be described by {\em quasiperiodic} functions via the ``projection method'', see e.g. \cite{deBruijn:1981:ATP,Duneau:1985:QP}. A function $f: \real^d \to \real$ is quasiperiodic if another function exists $g: \real^N \to \real$ with period $[0,2\pi]^N$ and a matrix $\mK \in \real^{d\times N}$ such that $d<N$, $f(\vx) = g(\mK^T \vx)$ and $\mK^T \vx = \vzero$ has only the trivial $\vx = \vzero$ solution when $\vx \in \real^d$ has entries with integer multiples of $2\pi$. Thus, $f$ is a restriction of the $N-$dimensional function $g$ to $d-$dimensions.

Let $p(\vx)$ be a scalar field describing a time-harmonic wave phenomenon, e.g. the acoustic pressure in a fluid with dispersed spherical particles. We model the interaction between the waves and particles using an energy landscape or potential $\psi(\vx) = U(p(\vx),\nabla p(\vx))$, whose minima correspond to locations where particles accumulate when subject to the field $p(\vx)$. This is a valid assumption for e.g. optics \cite{Ashkin:1987:OTM,Neuman:2004:OT,Nieminen:2007:POT} and ultrasound waves \cite{Friend:2011:MAM,Wiklund:2013:UEI,Courtney:2014:ITM,Greenhall:2015:UDS,Prisbrey:2017:UDS,Meng:2019:AT,Marzo:2019:HAT}. 
As we show in this paper, if $p(\vx)$ is a quasiperiodic function, then its energy landscape and corresponding pattern of particles must also be quasiperiodic. Moreover, a quasiperiodic $p(\vx)$ can result from a superposition of plane waves. Experimental evidence supporting this observation exists in e.g. optics, where five lasers (which can be modeled by plane waves) can create optical lattices with $10-$fold symmetries \cite{Roichman:2005:HAQ,Wang:2006:ROT,Mikhael:2008:ALT,Sun:2015:FTF}. 
Thus, the objective of this paper is to show a general theory to assemble patterns of particles dispersed in a fluid, for any linear wave phenomena. Further, we demonstrate the theory using two-dimensional ultrasound wave fields, established with $2N$ ultrasound transducers, where $N$ is the dimension of the higher dimensional space in the projection method \cite{deBruijn:1981:ATP,Duneau:1985:QP}. We disperse $80$ nm carbon nanoparticles in water and assemble them into quasiperiodic patterns with $8-$fold (octagonal) symmetry, using eight ultrasound transducers spatially arranged as a regular octagon. The theory is also valid in three dimensions. This theory is useful to conveniently fabricate materials with quasiperiodic patterns of particles embedded in a polymer matrix \cite{Greenhall:2015:UDS,Prisbrey:2017:UDS}, such as those used in engineered polymer composite materials and metamaterials, e.g. \cite{Corbitt:2015:IOD}.

%%%%%%%%%%%%%%%%%%%%%%%%%%%%%%%%%%%%%%%%%%%%%%%%%%%%%%%%%%%%%%%%%%%%%%%%
%\section{ARP and plane wave model}
The pressure associated with an ultrasound wave is given by $\widetilde{p}(\vx,t)= \Re( p(\vx) \exp[-i \omega t])$, where $\Re$ is the real part of a complex number, $\vx \in \real^d$, $t$ is time, $\omega$ is the angular frequency, and $i =\sqrt{-1}$. The complex valued field $p$ solves the  Helmholtz equation $\Delta p + k^2 p = 0$, with wavenumber $k = \omega /c$ and  wave propagation speed $c$. A small (relative to the wavelength) particle in a standing ultrasound wave is subject to the acoustic radiation force associated with that ultrasound wave \cite{Gorkov:1962:FAS, King:1934:ARP, Bruus:2012:ARF, Settnes:2012:FAS}. Thus, at location $\vx$ in an inviscid fluid, a small particle experiences a force $\vF(\vx) = - \grad \psi(\vx)$, where $\psi$ is the acoustic radiation potential (ARP). The ARP is given as
\begin{equation}
    \psi(\vx) = \mathfrak{a}|p(\vx)|^2 - \nabla p(\vx)^*\mathfrak{B} \nabla p(\vx).
    \label{eq:arp}
\end{equation}
Here $\mathfrak{a} = \mathfrak{f}_1 \kappa_0/4$,  $\mathfrak{B} = 3\mathfrak{f}_2 / (8\rho_0 \omega^2) \mI_d$, $\mI_d$ is the $d\times d$ identity matrix,  $\mathfrak{f}_1 = 1- (\kappa_p/\kappa_0)$, $\mathfrak{f}_2 = 2(\rho_p - \rho_0)/(2\rho_p + \rho_0)$ and $*$ is the conjugate transpose. The density and compressibility are $\rho$ and $\kappa$, with subscripts $0$ and $p$ referring to the fluid and particle, respectively. Particles assemble at the minima of the ARP because the acoustic radiation force vanishes where the ARP is (locally) minimum, and points towards the minimum in its vicinity. We remark that this theory neglects particle/particle interactions, i.e., it relies only on primary (direct) scattering. Furthermore the same theory describes the optical pressure exerted on dielectric particles that are smaller than a wavelength, by taking $\mathfrak{B} = \vzero$, see e.g. \cite{Neuman:2004:OT,Nieminen:2007:POT,Thomas:2017:AOR}.

\begin{figure}
    \centering
    \includegraphics[width = 0.20\textwidth]{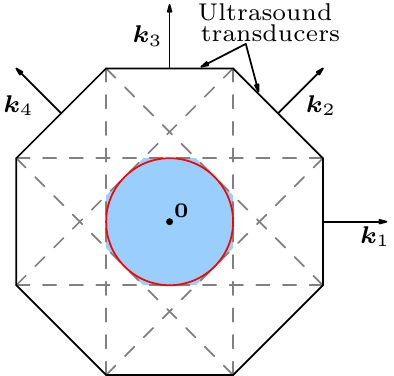}
    \caption{Each parallel pair of ultrasound transducers establishes an ultrasound wave that is close to a plane wave in the rectangle delineated by the dotted lines perpendicular to the ultrasound transducers. The plane wave model \eqref{eq:super} accurately represents the ultrasound wave field generated by this transducer arrangement in the blue region. The red circle indicates the region we evaluate.}
    \label{fig:region}
\end{figure}

We consider the particular case where the wave field $p(\vx)$ is a superposition of plane waves given by
\begin{equation}
p(\vx;\vu) = \sum\limits_{j=1}^N \alpha_j\exp[i\vk_j\cdot \vx]
+ \beta_j \exp[-i\vk_j\cdot\vx].
\label{eq:super}
\end{equation}
Here, $\vu = [\alpha_1,\ldots,\alpha_N,\beta_1,\ldots,\beta_N]^T$ is a vector containing the non-zero complex amplitudes in \eqref{eq:super} and $\vk_j$ are the wavevectors with $|\vk_j| = k$. We may obtain fields close to \eqref{eq:super} by using $N$ pairs of parallel-oriented ultrasound transducers with normal directions $\vk_j$ as is shown in \cref{fig:region} for $N=4$ and $d=2$. 
For each $j$, $\alpha_j$ and $\beta_j$ represent the amplitude and phase of the signals that drive a pair of parallel ultrasound transducers with normal $\vk_j$, as indicated in \cref{fig:region}. We call the vector $\vu$ ``transducer operating parameters'', although this ignores acoustic and electric impedances that would more accurately model the ultrasound transducers.  
The characterization of the quasiperiodic patterns of particles that can be achieved with the wave field \eqref{eq:super} is left for future studies. However, When $N=d$ and the $\vk_j$ are linearly independent in \eqref{eq:super}, the patterns of particles are periodic and are characterized in \cite{Guevara:2019:PPA}.

%%%%%%%%%%%%%%%%%%%%%%%%%%%%%%

%%%%%%%%%%%%%%%%%%%%%%%%%%%%%%%%%%%%%%%%%%%%%%%%%%%%%%%%%%%%%%%%%%%%%%%%
%\section{Quasiperiodicity}
The wave field \eqref{eq:super} is a restriction to dimension $d$ of a wave field in dimension $N$ with period $[0,2\pi]^N$. Thus, the wave field \eqref{eq:super} can be made quasiperiodic. To see this, we define $p_N(\vy;\vu)$ using \eqref{eq:super} with $\vy \in \real^N$ and the canonical basis vectors $\ve_j = (\delta_{ij})_{j=1}^N \in \real^N$ instead of the $\vk_j$, $j=1,\ldots,N$. Here we use $\delta_{ij} = 1$ if $i=j$ and $0$ otherwise. Clearly $p_N(\vy;\vu)$ is periodic in $\vy$ for any choice of complex amplitudes $\vu$, and its period is the hypercube $[0,2\pi]^N$. We use the convention that $\vx \in \real^d$ and $\vy \in \real^N$. A simple calculation reveals that
\begin{equation}
    p(\vx;\vu) = p_N(\mK^T \vx;\vu),
    \label{eq:pn}
\end{equation}
where $\mK = [\vk_1,\ldots,\vk_N] \in \real^{d \times N}$ determines whether the wave field is periodic or quasiperiodic. Naturally, the ARP $\psi$ in dimension $d$ (see \eqref{eq:arp}) relates to a similar quantity $\psi_N$ in dimension $N$ that is of the same form, but involves $p_N$ instead of $p$ and with identical $\mathfrak{a}$ but where the matrix $\mathfrak{B}_N = \mK^T \mathfrak{B} \mK \in \real^{N \times N}$ is different because of the chain rule, i.e.
\begin{equation}
 \psi(\vx;\vu) = \psi_N(\mK^T \vx; \vu).
 \label{eq:psin}
\end{equation}
Hence, the ARP is quasiperiodic in $\vx$ if $p(\vx;\vu)$ is quasiperiodic.

\begin{figure}
    \centering
    \begin{tabular}{ccc@{\hspace{0em}}c}
     \includegraphics[width = 0.14\textwidth]{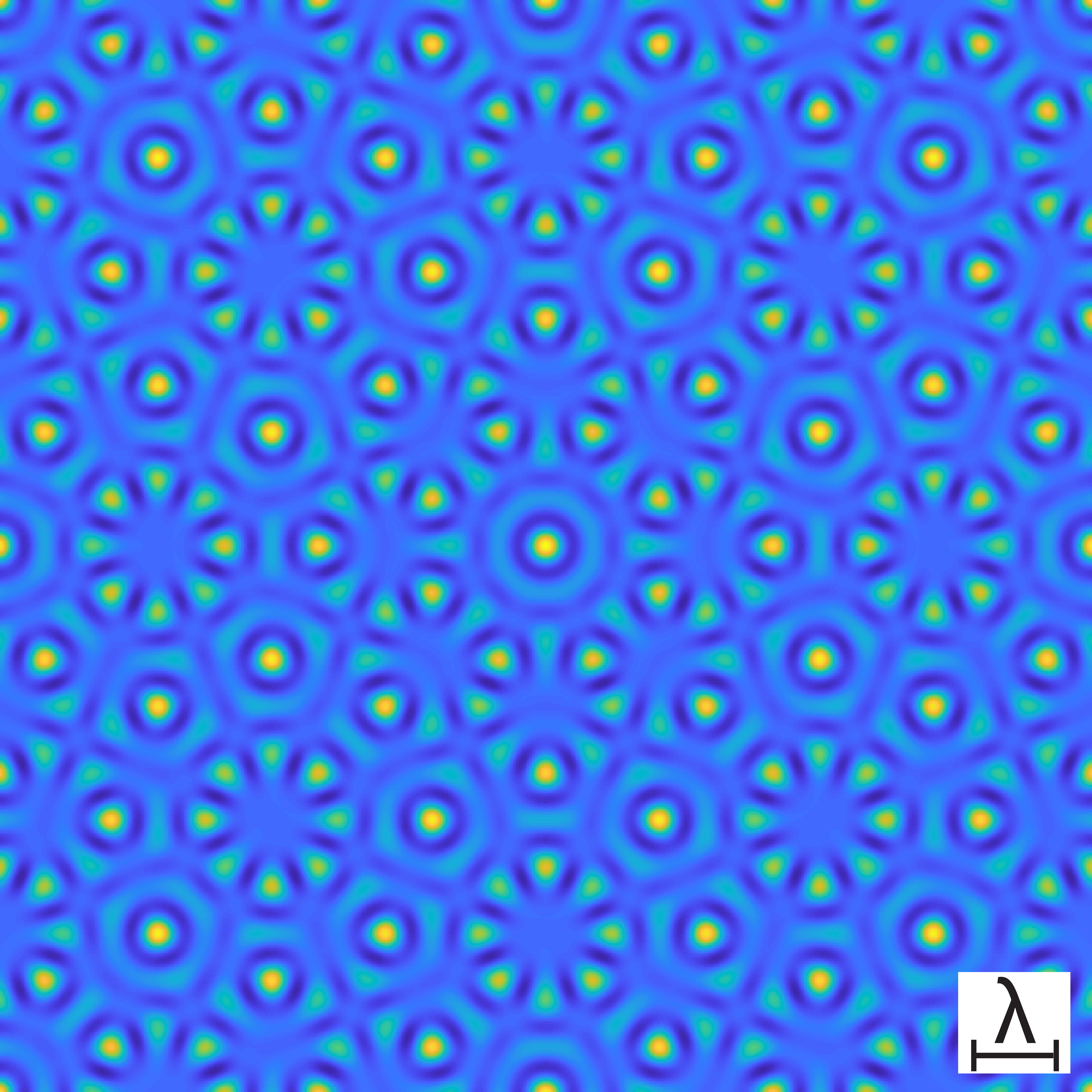} & 
     \includegraphics[width = 0.14\textwidth]{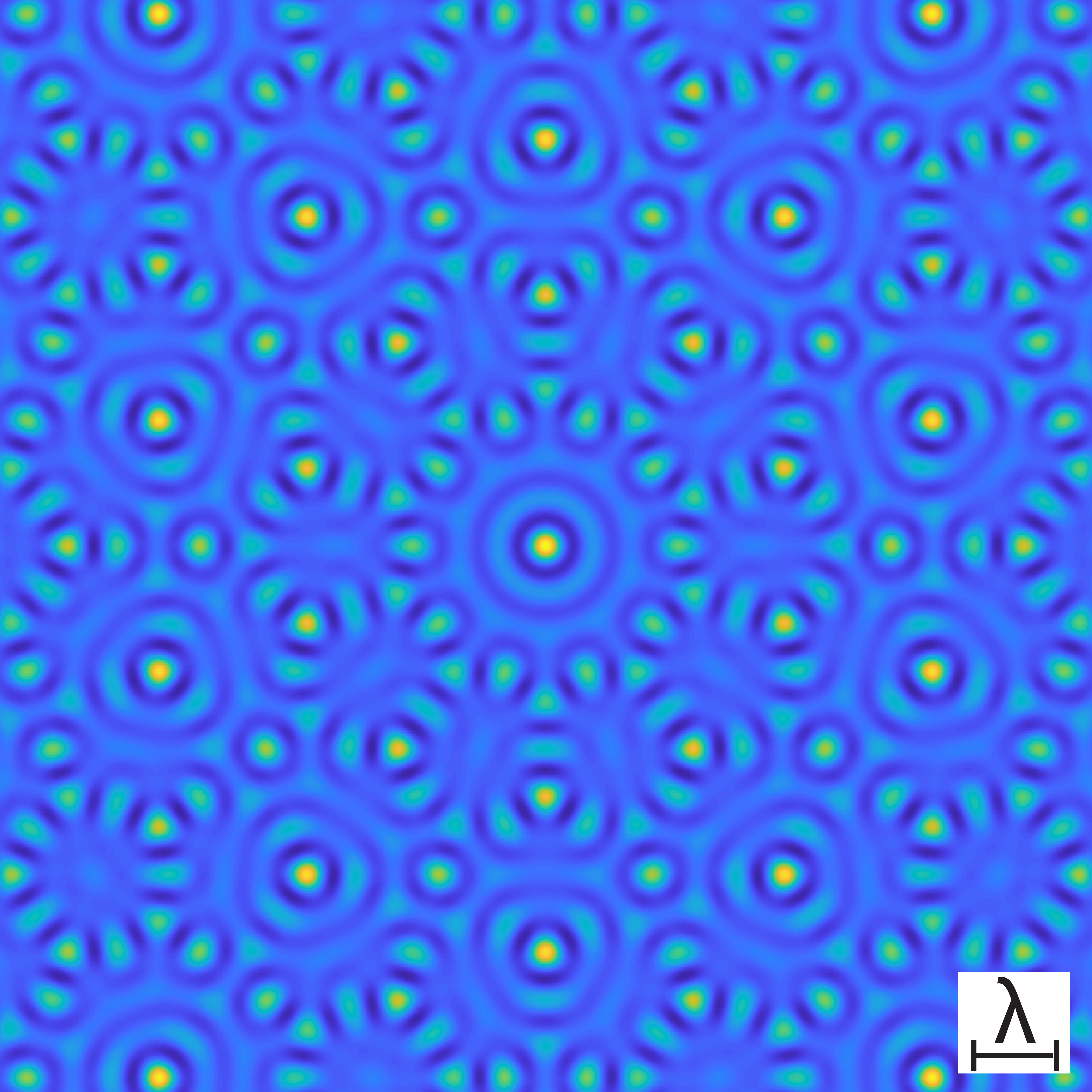} &
     \includegraphics[width = 0.14\textwidth]{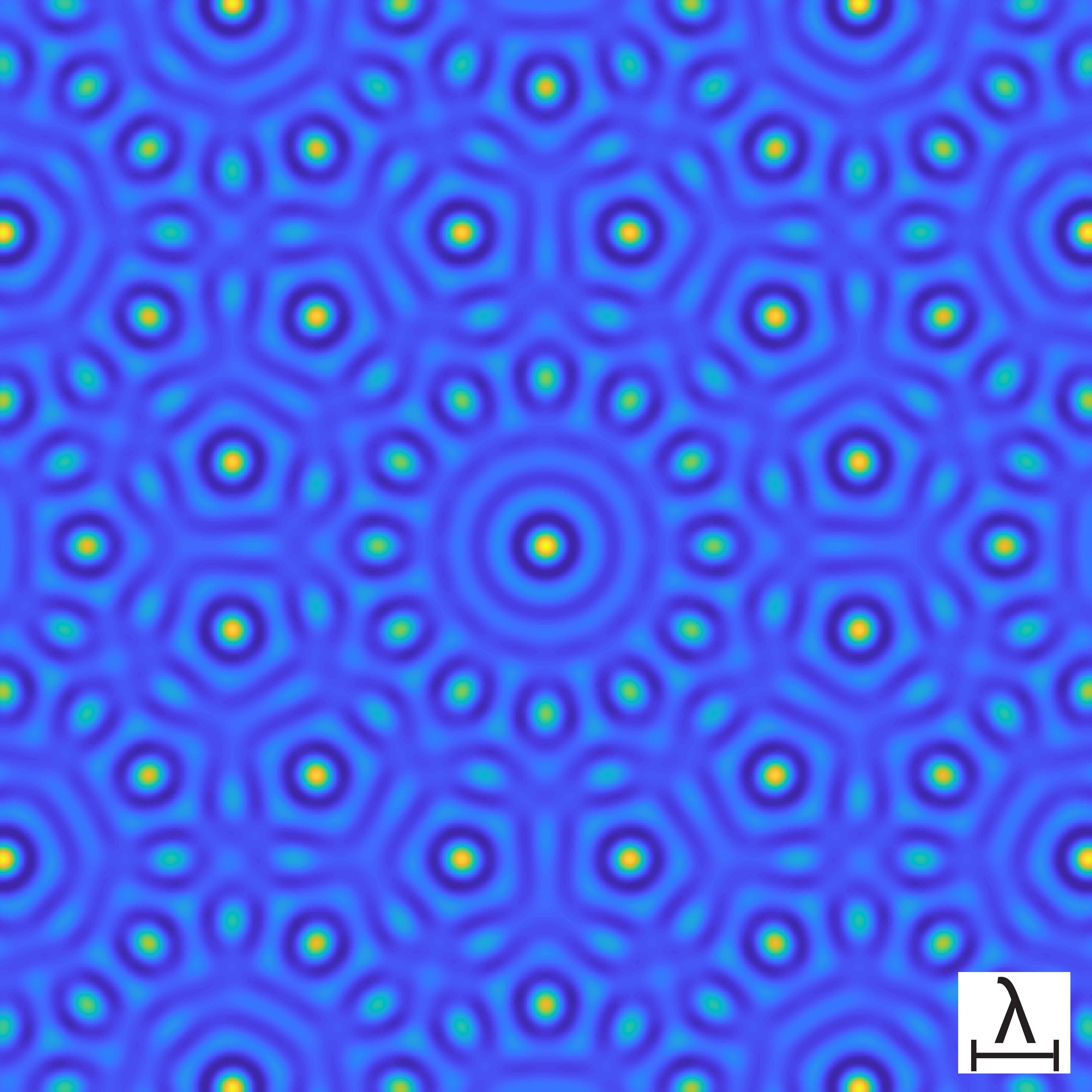} & 
     \raisebox{-0.5em}{\includegraphics[width = 0.050\textwidth]{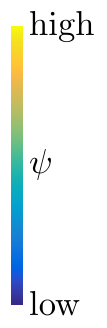}}\\
     $8-$fold symmetry &  $10-$fold symmetry & $12-$fold symmetry
    \end{tabular}
    \caption{Examples of ARP that lead to quasiperiodic two-dimensional patterns of particles. The wavevectors are $\vk_j = [\cos\theta_j,\sin\theta_j]^T$, where $\theta_j = j\pi/N$, $j=0,\ldots,N-1$ and $2N \in \{8,10,12\}$ corresponds to the desired order of rotational symmetry, i.e., $8-$, $10-$ or $12-$fold symmetry. The ultrasound transducer operating parameters in \eqref{eq:super} are $\vu = [1,\ldots,1]^T \in \real^{2N}$. The computation was performed using a uniform grid of the square $[-7\lambda,7\lambda]^2$ with $1024^2$ points. The color scale shows arbitrary units.}
    \label{fig:theo}
\end{figure}

The superposition of plane waves \eqref{eq:super} predicts two-dimensional quasiperiodic patterns of particles with prescribed symmetries. In the particular case of $80$ nm carbon nanoparticles dispersed in water, we use $c_0 = 1500$ m/s, $c_p = 5300$ m/s, $\rho_0 = 1000$ kg/m$^3$ and $\rho_p = 2100$ kg/m$^3$. Since $\kappa = 1/(\rho c^2)$, we obtain $\mathfrak{a} \approx 5.7424 \times 10^6$ and $\mathfrak{B} \approx (0.2115) \mI_2$ in \eqref{eq:arp}. We intend the patterns of particles within the octagonal arrangement of ultrasound transducers in \cref{fig:region} to show $8-$fold symmetry. \Cref{fig:theo} illustrates simulations of different symmetries for the carbon nanoparticles in water in the far field, i.e., the ultrasound transducers are sufficiently far from the region that we display. For instance, if the $8$ ultrasound transducers in \cref{fig:region} are driven with the same amplitude and phase, i.e., $\vu = [1,\ldots,1]^T \in \real^{8}$, we obtain patterns of particles with an $8-$fold symmetry and center of rotation at the origin (which in \cref{fig:region} corresponds to the center of the red circle). \Cref{fig:theo} also shows quasiperiodic patterns of particles with $10-$fold and $12-$fold symmetries that can be obtained by arranging the ultrasound transducers in \cref{fig:region} as either a regular decagon or dodecagon, instead of an octagon. Thus, we obtain known quasi-periodic symmetries in two dimensions, e.g. the $10-$fold symmetry is the same symmetry encountered in Penrose tilings \cite{Penrose:1979:PCN,deBruijn:1981:ATP,Yamamoto:1996:CQC}. We also point out that ultrasound transducer arrangements other than regular polygons are possible and will yield patterns of particles with other symmetries.
    
%%%%%%%%%%%%%%%%%%%%%%%%%%%%%%

%%%%%%%%%%%%%%%%%%%%%%%%%%%%%%%%%%%%%%%%%%%%%%%%%%%%%%%%%%%%%%%%%%%%%%%%
%\section{Experimental setup}

\begin{figure}
    \centering
    \begin{tabular}{cc}
     \includegraphics[width = 0.24\textwidth]{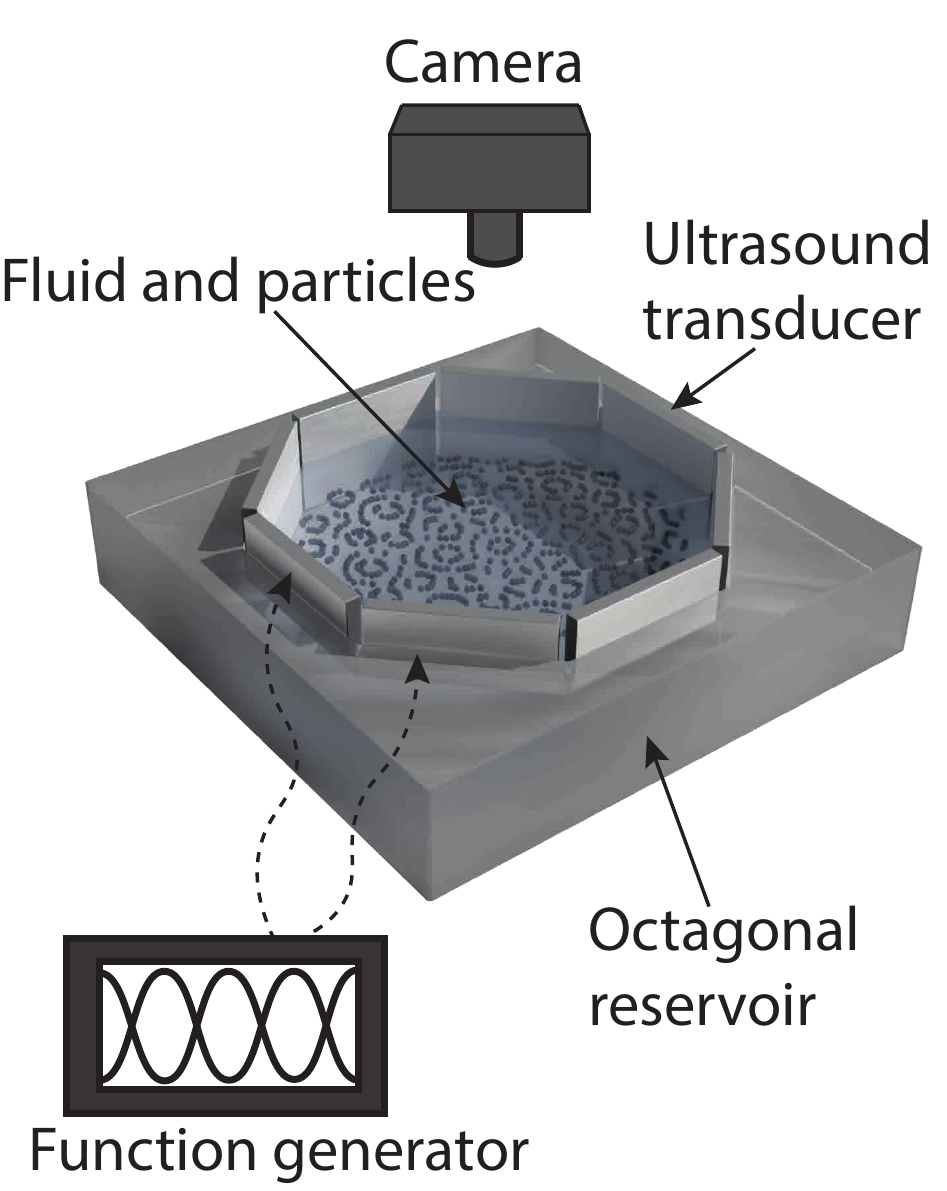}
      & \includegraphics[width = 0.22\textwidth]{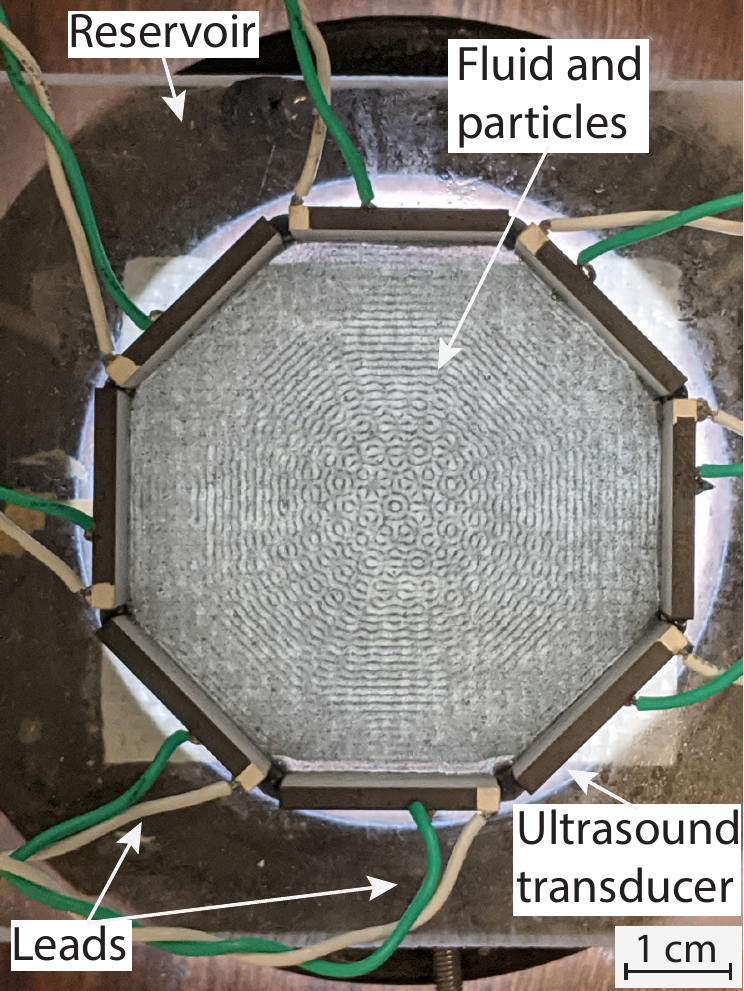}\\
     (a) & (b)
    \end{tabular}
    \caption{(a) Schematic of the experimental setup (isometric view) showing its different components and (b) photograph of the experimental setup (top view), showing a typical experiment with 80 nm carbon nanoparticles dispersed in water.}
    \label{fig:schema}
\end{figure}

We used the setup of \cref{fig:schema} to experimentally obtain quasiperiodic patterns with $8-$fold symmetries. The setup comprises a polycarbonate octagonal reservoir with water, 80 nm carbon nanoparticles, and sodium dodecyl benzene sulfonate (NaDDBS) surfactant \cite{Islam:2003:HWF}, and it is lined with 8 ultrasound transducers along its perimeter (SM111 piezoelectric material, with center frequency of 1 MHz), which are driven by a function generator. The distance between two parallel ultrasound transducer is 5 cm and each ultrasound transducer has a width of $2$ cm (or $40\lambda/3$).

%%%%%%%%%%%%%%%%%%%%%%%%%%%%%%

\begin{figure}
    \centering
    \begin{tabular}{cccc}
    & Simulated ARP & Experiment photo & Comparison \\
    \raisebox{1em}{\rotatebox{90}{Experiment 1}} & 
    \includegraphics[width = 0.15\textwidth]{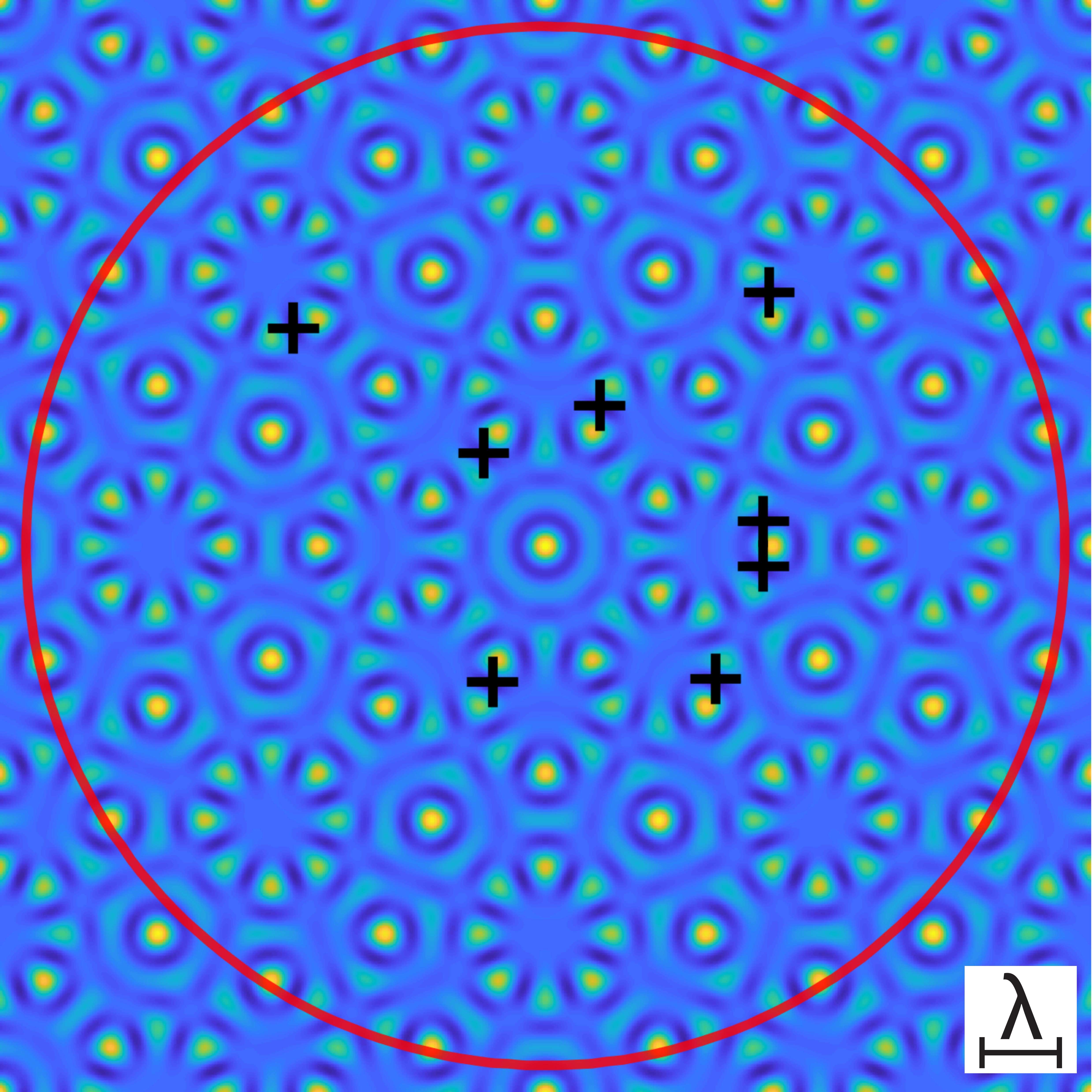}& 
    \includegraphics[width = 0.15\textwidth]{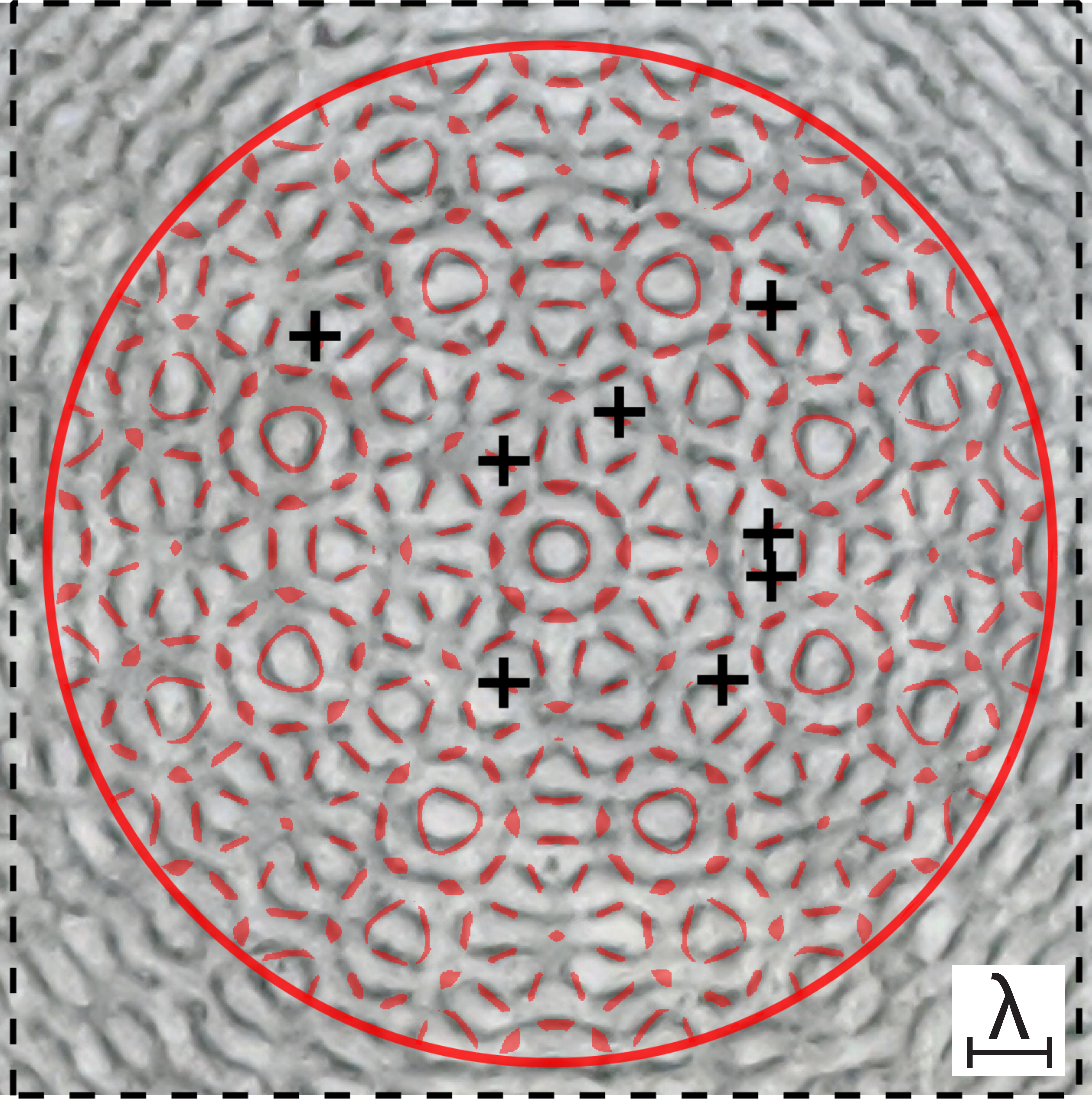} & \includegraphics[width = 0.15\textwidth]{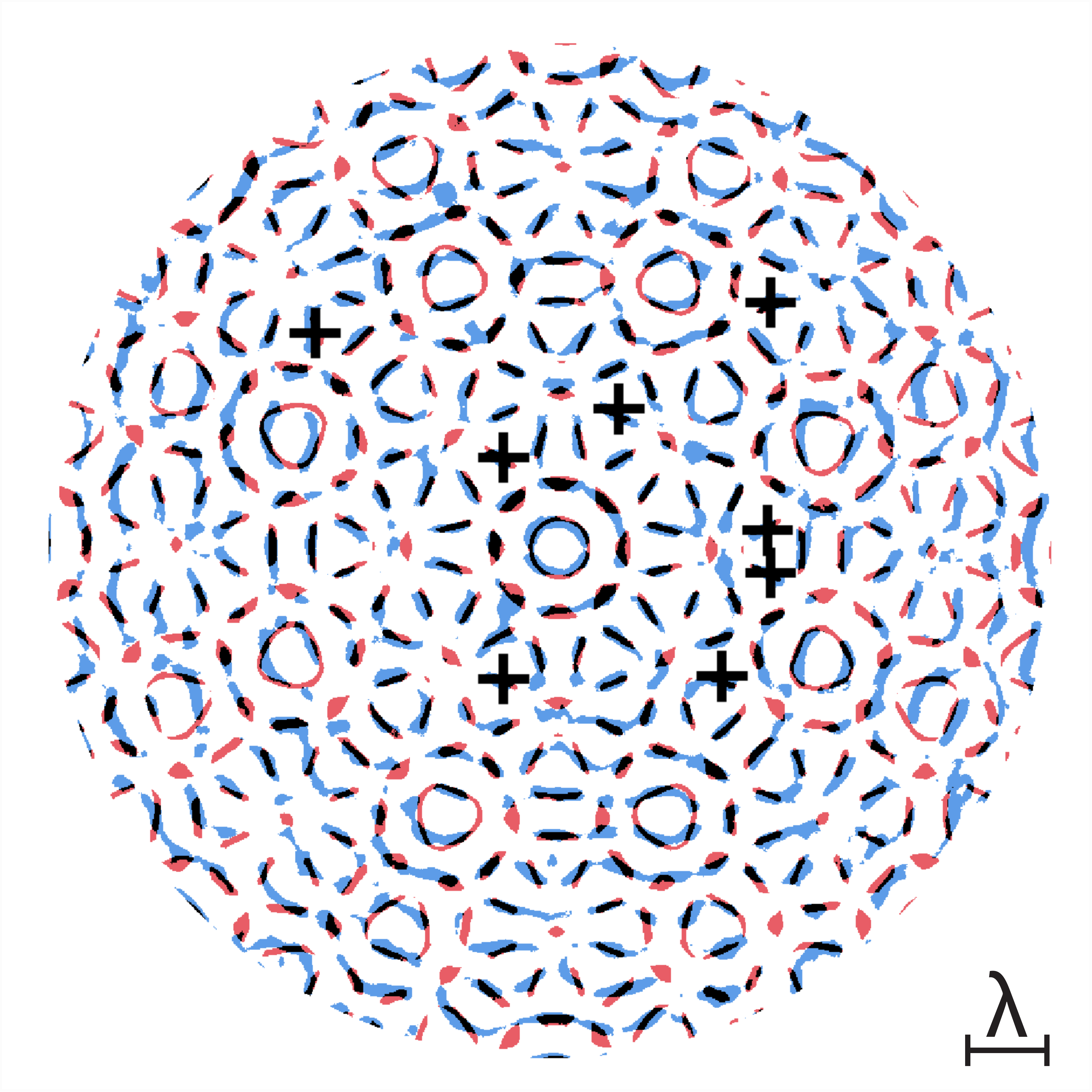} \\
    \raisebox{1em}{\rotatebox{90}{Experiment 2}} & 
    \includegraphics[width = 0.15\textwidth]{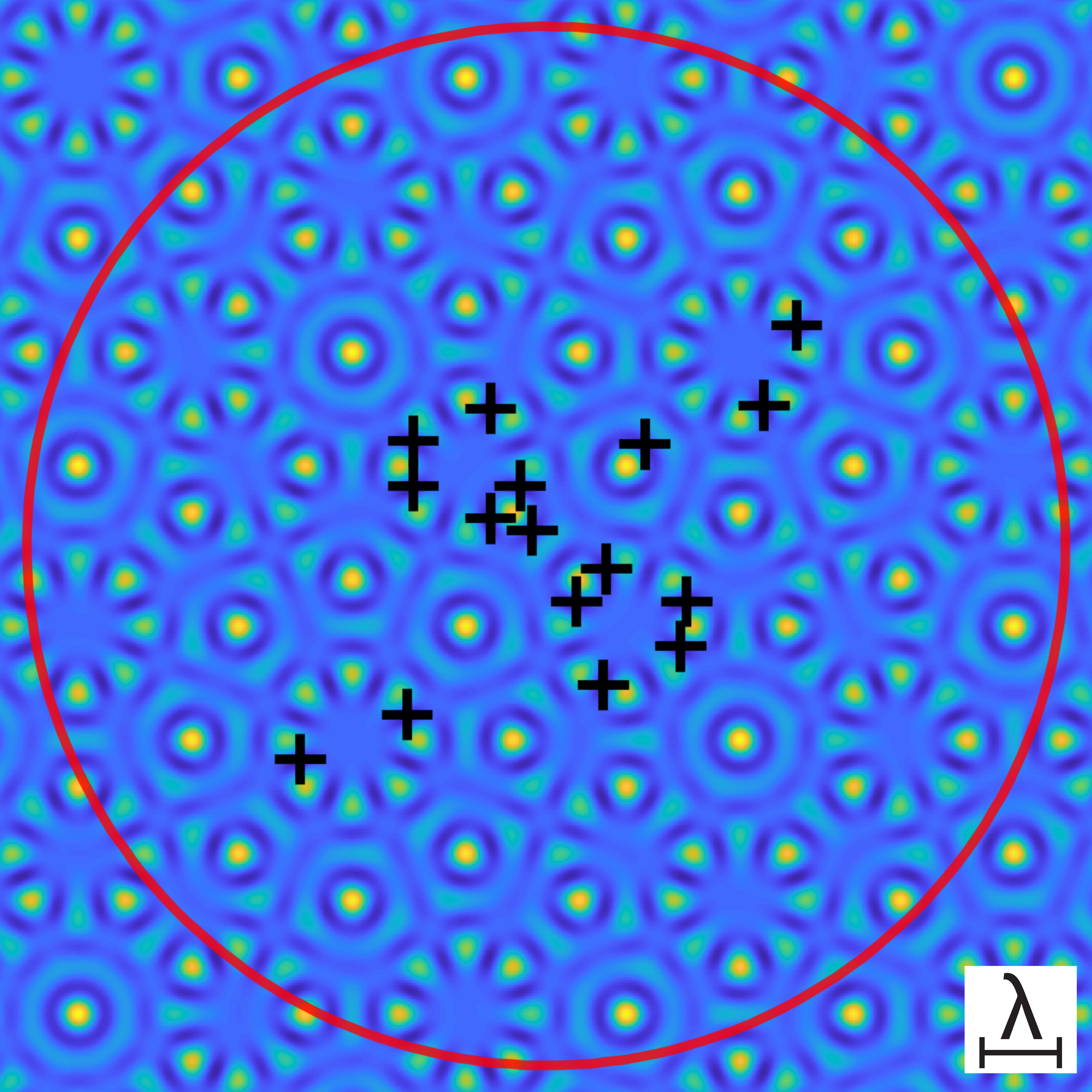} &
    \includegraphics[width = 0.15\textwidth]{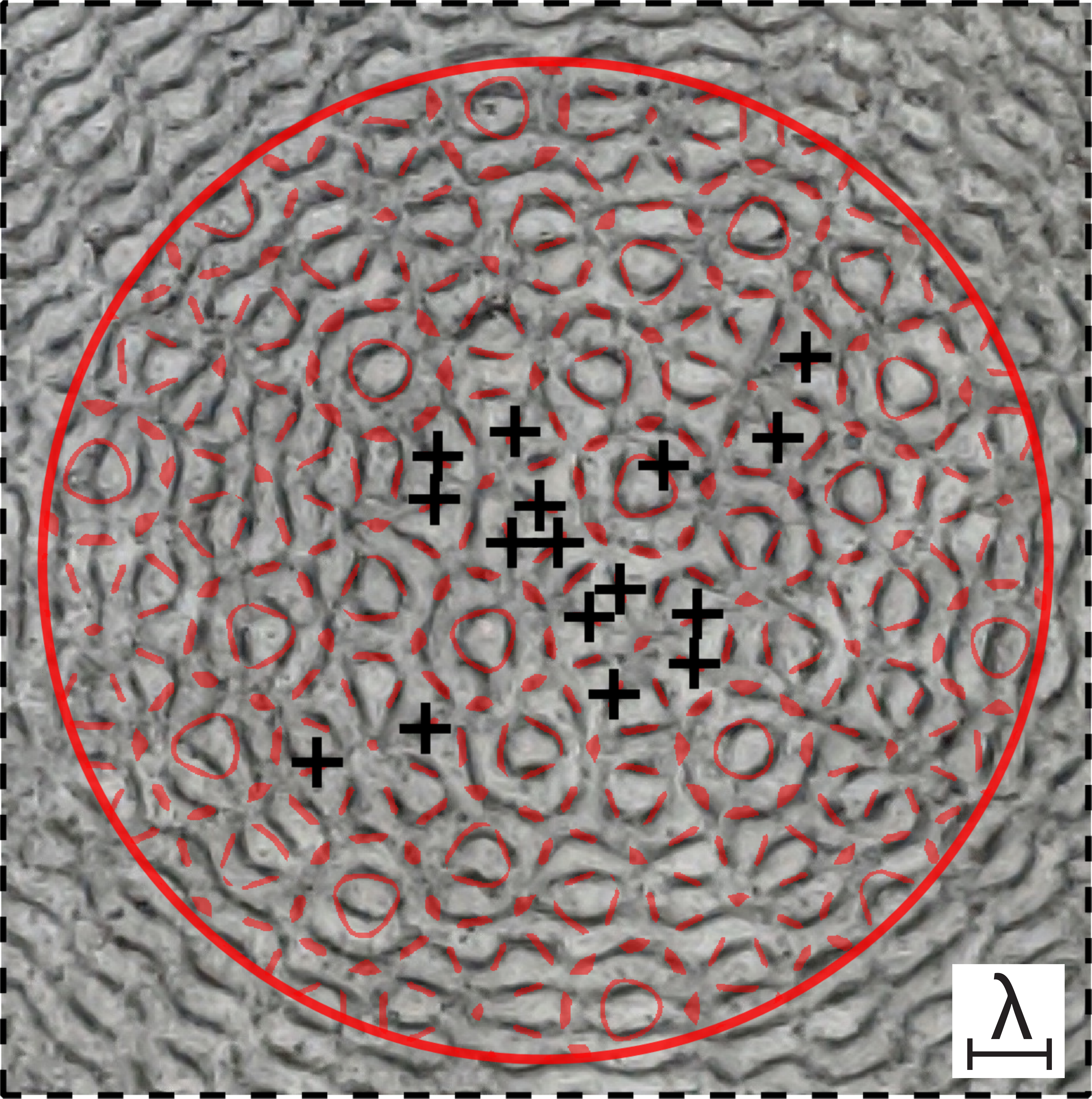} &
    \includegraphics[width = 0.15\textwidth]{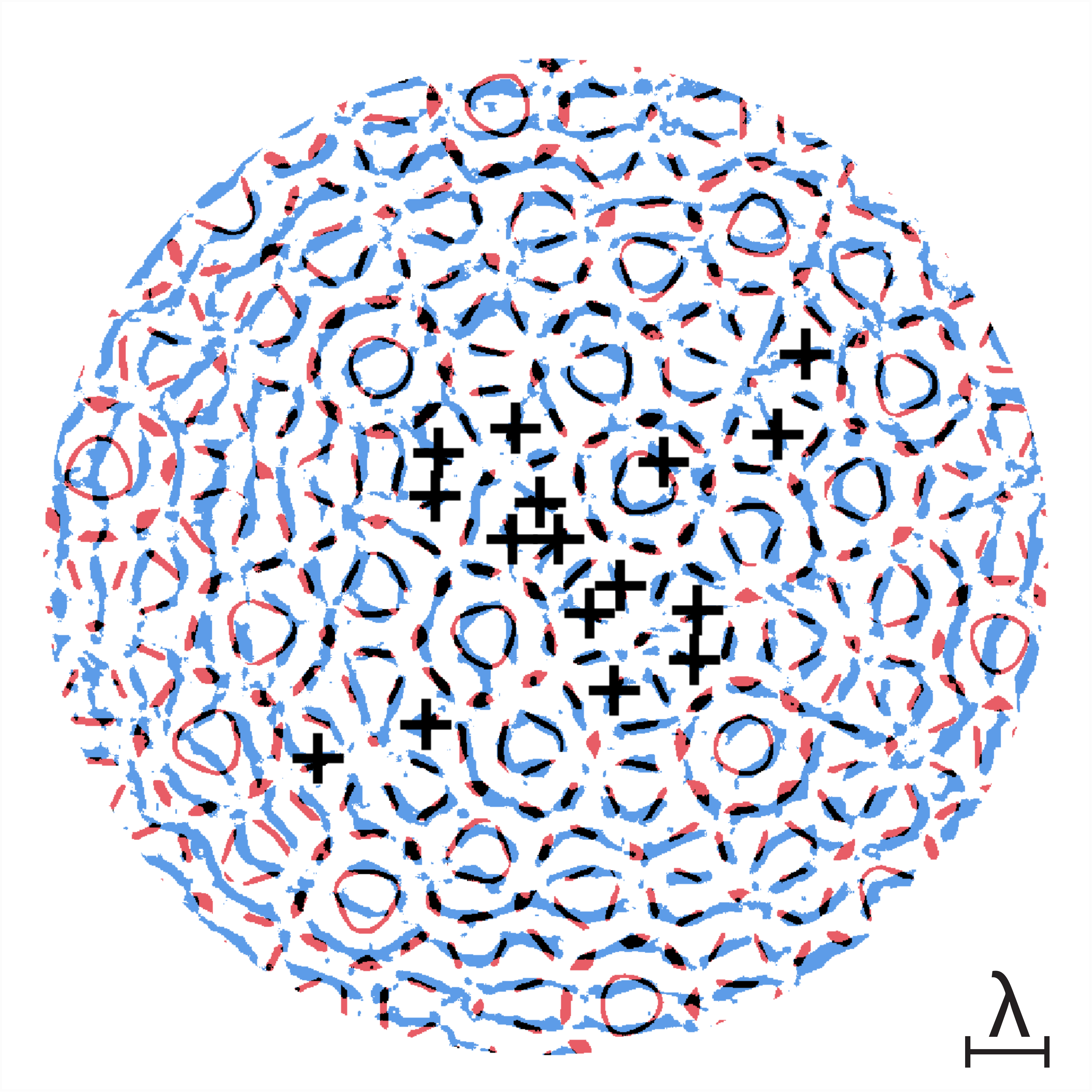}\\[-0.2em]
    &\includegraphics[width = 0.16\textwidth]{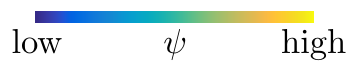} &&
    \end{tabular}
    \caption{Simulated ARP (first column, particles assemble at minima) and experimentally obtained patterns of particles with the minima of the simulated ARP superimposed in red (second column), for Experiments 1 and 2, showing distinct quasi-periodic patterns with $8-$fold symmetries. We superimpose the simulated ARP minima where we predict particles assemble (red) and experimentally obtained patterns of particles (blue) (third column). We mark the overlapping locations (blue and red) in black. Each image shows a red circle where we expect good agreement with the plane wave model (see also \cref{fig:region}) and ``plus signs'' indicate the registration between the simulated ARP and the experimental results.}
    \label{fig:exp}
\end{figure}

We compare simulations of the ARP obtained from the theoretical plane wave superposition \eqref{eq:super}, where particles assemble at the minima of the ARP, to experimentally obtained patterns of particles, for two specific experiments. 
In experiment 1, we impose ultrasound transducer operating parameters $\vu = [1,1,1,1,1,1,1,1]^T$ and in experiment 2  we impose $\vu = [1,-1,1,1,1,-1,1,1]^T$. \Cref{fig:exp} shows the simulated ARP in the region $[-7\lambda, 7\lambda]^2$, and we indicate the region we evaluate as a red circle, as in \cref{fig:region}. We used explicit expressions for the gradient and Hessian of the ARP to predict the locations where particles assemble, by identifying points where the Hessian is sufficiently positive definite (minimum eigenvalue greater than $10^{-6})$ and the gradient is sufficiently small (less than $4\times 10^{11}$). We show the simulated locations where particles assemble in red, superimposed on the experimental results (photographs). We manually registered the simulated and experimental results with Matlab's {\tt\small fitgeotrans} at the points indicated by plus signs in \cref{fig:exp}, assuming a 2D projective transformation. Finally, we qualitatively compare the simulated and experimental results by superimposing the simulated locations where we predict particles assemble (red) and experimentally obtained patterns of particles (blue), and we mark the overlapping locations (blue and red) in black. We observe good qualitative agreement between simulations and experiments.

%%%%%%%%%%%%%%%%%%%%%%%%%%%%%%

\begin{figure}
    \centering
    \begin{tabular}{cc}
    \rotatebox{90}{\parbox{5cm}{Agreement between simulations \\and experiments [\%]}}
    &
    \includegraphics[width = 0.35\textwidth]{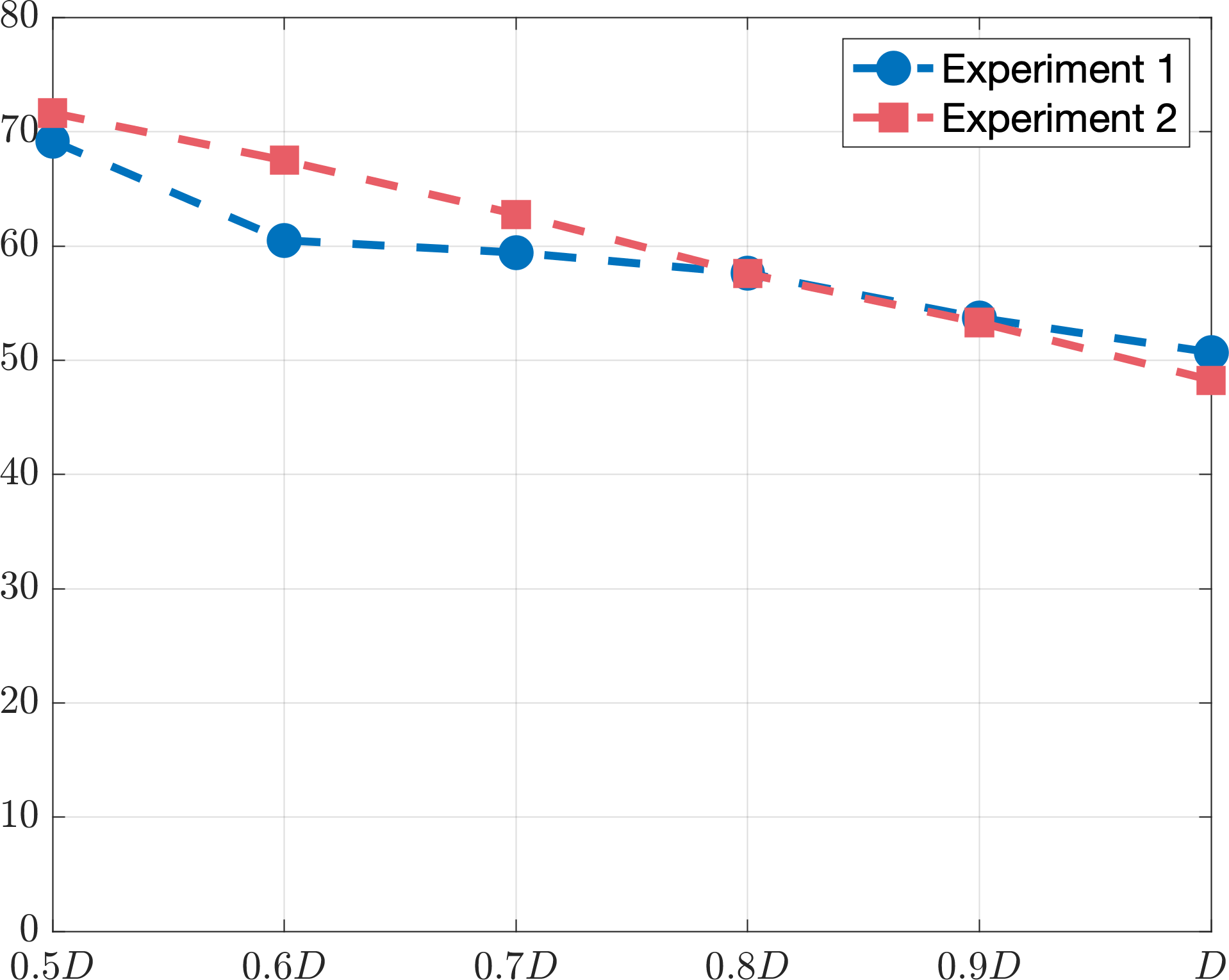} \\
    & Diameter of evaluation circle [$D$]
    \end{tabular}
    \caption{Agreement fraction (in percentage) between simulations and experiments as a function of evaluation circle diameter. Here $D = 40\lambda/3$ is the transducer width.}
    \label{fig:disks}
\end{figure}

We quantified the agreement between simulations and experiments within the red circle of diameter $D=40\lambda/3$ as follows. We binarized the photographs of the experimental results, using Matlab's {\tt\small imbinarize} with sensitivity 0.45. We determined the fraction (in percentage) of the total area of the simulated clusters that is inside the experimentally determined clusters, i.e., (black area)/(black + blue area) using the color convention in the comparison column of \cref{fig:exp}. \Cref{fig:disks} shows that the agreement between simulations and experiments improves with decreasing size of the evaluation circle with diameter $\alpha D$, with $\alpha \in [1/2,1]$. We observe from \cref{fig:disks} that the fraction of agreement between simulations and experiments increases linearly with decreasing size of the evaluation circle. Thus, the simulations and experiments agree most closely at the origin, which is consistent with \eqref{eq:super} being a far field model. Other sources of error in the model \eqref{eq:super} are that it neglects the boundary reflections and the finite width of the ultrasound transducers.

%%%%%%%%%%%%%%%%%%%%%%%%%%%%%%

%%%%%%%%%%%%%%%%%%%%%%%%%%%%%%%%%%%%%%%%%%%%%%%%%%%%%%%%%%%%%%%%%%%%%%%%
%\section{Summary and perspectives}
We have shown that linear wave phenomena can assemble quasiperiodic patterns of particles. Since linear wave phenomena are common in physics, our findings apply to a variety of different physical situations, including ultrasound waves, electromagnetic waves, elastic waves, amongst others. We illustrated this principle theoretically with plane waves in a fluid, and demonstrated it experimentally by assembling 80 nm carbon nanoparticles dispersed in water into patterns with $8-$fold symmetries, using ultrasound waves. The theory accurately predicts the experimental patterns of particles. Thus, this theory and experimental demonstration provides a pathway upon which to base a manufacturing platform for quasicrystal-like structures of inclusions in a polymer matrix. Such materials could help with the experimental study of the physical properties of quasicrystals. These quasiperiodic materials could have different mechanical or electrical properties than a material with a random arrangement of particles. 
%To complete the theoretical understanding of this method for obtaining quasiperiodic materials, we are working on an explicit characterization of the minima of the ARP that relies on the geometry of the projection from a higher dimensional space and the minima of the ARP-like quantity in higher dimensions, which can be found explicitly.

\begin{acknowledgments}
 F.G.V., C.M., M.P. and B.R. acknowledge support from the Army Research Office under contract No. W911NF-16-1-0457.
\end{acknowledgments}

The mathematical theory and results were derived by E.C., F.G.V. and C.M. The simulations were designed and performed by F.G.V., C.M. and M.P. using Matlab. The experiments were designed by all authors and implemented and performed by M.P. and B.R. All authors contributed to writing the manuscript.

%%%%%%%%%%%%%%%%%%%%%%%%%%%%%%%%%%%%%%%%%%%%%%%%%%%%%%%%%%%%%%%%%%%%%%%%

%\bibliographystyle{abbrv} % revtex picks this up automatically
\bibliography{herglotz_bib}
\end{document}